\documentstyle[preprint,aps]{revtex}

\begin{document}
\draft


\preprint{IUNTC \#93--07}
\title{Role of heavy-meson exchange in pion production near threshold}
\author{C.~J. Horowitz,$^{\rm (1)}$%
\thanks{Electonic communication to charlie@venus.iucf.indiana.edu}
H.~O. Meyer,$^{\rm (2)}$ and David K. Griegel$^{\rm (1)}$}
\address{$^{\rm (1)}$Department of Physics and Nuclear Theory Center\\
Indiana University, Bloomington, Indiana 47405}
\address{$^{\rm (2)}$Department of Physics and Cyclotron Facility\\
Indiana University, Bloomington, Indiana 47405}
\date{April 1993}
\maketitle
\begin{abstract}
Recent calculations of $s$-wave pion production have severely
underestimated the accurately known $pp\rightarrow pp\pi^0$\ total
cross section near threshold.
In these calculations, only the single-nucleon axial-charge operator
is considered.
We have calculated, in addition to the one-body term, the two-body
contributions to this reaction that arise from the exchange of mesons.
We find that the inclusion of the scalar $\sigma$-meson exchange
current (and lesser contributions from other mesons) increases the
cross section by about a factor of five, and leads to excellent
agreement with the data.
The results are neither very sensitive to changes in the distorting
potential that generates the $NN$ wave function, nor to different
choices for the meson-nucleon form factors.
We argue that $pp\rightarrow pp\pi^0$\ data provide direct
experimental evidence for meson-exchange contributions to the axial
current.
\end{abstract}


\section{INTRODUCTION}
\label{sec:introduction}

Recently, a precise measurement of the total cross section for pion
production in the reaction $pp\rightarrow pp\pi^0$\ was carried out
using the electron-cooled stored beam of the Indiana University
Cyclotron Facility (IUCF) Cooler with an internal gas target \cite{MEY92}.
The novel technology made it possible to extend the measurements to
within a few MeV of threshold.
Over most of the covered energy range, contributions from higher
partial waves are negligible, and the cross section is thus due to a
single partial wave ($^3P_0\rightarrow{}^1S_0+s$-wave pion).
That this is the case can be deduced from the energy dependence of the
cross section and the angular distribution of the outgoing protons
\cite{MEY92}.

As we will discuss below, $s$-wave pion production is sensitive to the
axial charge of the two-nucleon system.
A number of calculations using the single-nucleon axial-charge operator
\cite{KOL66,MIL91,NIS92,MEY92} have been carried out with various choices for
the $NN$ distorting potentials.
Quite surprisingly, such calculations all underestimate the data by
approximately a factor of five.
This discrepancy is even more serious since there is little ambiguity
in the calculations, because only a single partial wave is involved,
and the $NN$ wave function and the (free) operator are both well
known.

The discrepancy could be explained by an enhancement of the axial
charge in the $NN$ system, which could come from a relativistic effect.
There exist many relativistic calculations for nuclear systems; for
example, relativistic impulse-approximation calculations \cite{SHE83,MUR87}
accurately reproduce elastic proton-nucleus scattering data.
Relativistic models characteristically feature large Lorentz
scalar and vector potentials \cite{CLA83}.
The strong scalar potential enhances the lower component of the Dirac wave
function of the nucleon; this change enhances the axial charge \cite{MCN85}
(see Sec.~\ref{sec:implications}).
Several authors have examined the consequences of this relativistic
effect on $\beta$ decay or muon capture rates in $A=16$ nuclei \cite{DOD85}
and nuclei near $A=208$ \cite{GAT92}.

In nonrelativistic models, relativistic effects can be incorporated
formally via meson-exchange currents (MEC's).
In this view the relativistic effect would be represented by a MEC
involving  a scalar $\sigma$\ meson.
Note that $\sigma$\ exchange also provides a phenomenological model
for the intermediate-range attraction in the $NN$ interaction and is
responsible for part of the spin-orbit potential.
Many authors have examined MEC contributions to the axial-charge
operator in nuclei, and found that the largest contribution is due to
$\pi\rho$ exchange \cite{TOW92}, which enhances axial-charge matrix
elements by about 60\%.
The next largest MEC is believed to be the relativistic
$\sigma$\ exchange, contributing another 40\% to the axial charge.
Together these contributions could account for the approximately 100\%
enhancement seen in axial-charge matrix elements for a variety of
$\beta$ decays \cite{TOW92}.
However, the size of the MEC contribution that one would derive
from experiment depends on assumptions about complicated wave
functions and short-range $NN$ correlations.

Because of isospin considerations, the $\pi\rho$ current (proportional
to the dot product of the isospin matrices
$\vec{\tau}_1\times\vec{\tau}_2$\ and the pion field) cannot
contribute to $pp\rightarrow pp\pi^0$.
However, other MEC may well be important.
In this case, pion production in the $pp$ system would provide a
unique laboratory for testing MEC models for a number of reasons.
First, the initial proton must decelerate in order to produce a pion.
The resulting large momentum mismatch can best be mediated by meson
exchange.
Heavy mesons are favored; indeed, the initial relative momentum near
threshold of about $1.9\nobreak\,\text{fm}^{-1}$\ is  comparable to the mass of
the $\sigma$\ meson.
Thus the process is very sensitive to two-body contributions.
Second, contributions from intermediate states with a nucleon and a
spin-$3\over 2$ $\Delta$ are expected to be small since such a system
cannot be formed in
a relative $s$ state because of angular momentum considerations.
Furthermore, the $\Delta$ cannot decay into a spin-$1\over 2$ nucleon and
$s$-wave pion.
(Reference~\cite{NIS92} provides estimates of $\Delta$ contributions.)
Third, pion rescattering effects are small because the scattering length for
pion-nucleon $s$-wave scattering in the required isospin combination is small,
as will be discussed in Sec.~\ref{sec:formalism}.
Finally, the wave functions in this case are simple and can be
calculated reliably.

Lee and Riska \cite{LEE93} have recently suggested that the inclusion
of MEC could explain the magnitude of the observed
$pp\rightarrow pp\pi^0$\ total cross section.
They calculate MEC by assuming a simple operator form for the $NN$ potential.
This allows the calculation of MEC for phenomenological potentials, but depends
on model assumptions.
In this paper, we perform a similar calculation; however, we use an
explicit one-boson-exchange model for the $NN$ interaction as well as
for the calculation of meson-exchange contributions.
We also include the Coulomb interaction, which is quite important near
threshold.
Furthermore, we examine the sensitivity of the calculated cross
section to many of the model ingredients.

In Sec.~\ref{sec:formalism}, we outline the formalism for the
calculation and discuss the input parameters and some computational
details.
In Sec.~\ref{sec:discussion} we collect and discuss the results for a
variety of one-body and two-body contributions and for different $NN$
interactions.
Section~\ref{sec:implications} lists the conclusions from our work.


\section{FORMALISM AND DETAILS OF THE CALCULATION}
\label{sec:formalism}

In this section we present the formalism used to calculate the total
$pp\rightarrow pp\pi^0$\ cross section for $s$-wave pion production.
The calculation is carried out in coordinate space because this allows
for a simple treatment of the Coulomb interaction between the two
protons.
The first three subsections are devoted to the matrix elements that
correspond to the production mechanisms illustrated by the diagrams in
Fig.~\ref{fig1}, namely, (a) the one-body term and the two-body terms
that arise either from (b) pion rescattering or (c) the exchange of
heavier mesons.
In Sec.~\ref{sub:calculational}, we collect a number of calculational
details and give the expression for the total cross section.


\subsection{One-body matrix element}
\label{sub:one_body}

The one-body term [Fig.~\ref{fig1}(a)] can be viewed as a nucleon
radiating a pion in the distorting potential of the other nucleon.
The interaction between the two nucleons is needed in order to
conserve energy and momentum.

We start with the pseudovector interaction Lagrangian between the pion
field $\vec{\phi}$ and the nucleon field $\psi$ \cite{WEI67},
\begin{equation}
{\cal L_{\rm int}}=-{f_{\pi NN}\over m_\pi}
\overline{\psi}\gamma_5\gamma^\mu\vec{\tau}\psi
\cdot\partial_\mu\vec{\phi}\ ,
\label{eq:Lint}
\end{equation}
where $f_{\pi NN}$\ is the pion coupling constant and $m_\pi$\ is the
pion mass.
A {\it pseudoscalar\/} Lagrangian (proportional to
$\overline{\psi}\gamma_5\vec{\tau}\psi\cdot\vec{\phi}$), instead of
Eq.~(\ref{eq:Lint}), would yield the same one-body contribution;
however, {\it pseudovector\/} coupling is favored for reasons that are
discussed in Sec.~\ref{sub:exchange}.

The field operators in Eq.~(\ref{eq:Lint}) involve sums over operators
that create or destroy particles of a given momentum \cite{BJO64}.
The term
\begin{equation}
{\cal M}_{fi}\propto\overline{U}_{s^\prime}(p^\prime)(\gamma_5\gamma_0 q_0
-\gamma_5\bbox{\gamma}\cdot{\bf q})U_s(p)\ ,
\label{eq:emit}
\end{equation}
is responsible for the emission of a neutral pion of momentum $q$\ by a
proton of momentum $p$.
Here $U$\ is a  proton Dirac spinor, and $p^\prime=p-q$\ is the
final proton momentum.
The second term in Eq.~(\ref{eq:emit}) has the nonrelativistic limit
$\gamma_5\bbox{\gamma}\rightarrow -\bbox{\sigma}$, where
$\bbox{\sigma}$\ stands for the conventional  $2\times 2$\ Pauli
spin matrices.
This term, which represents the familiar nonrelativistic
$\bbox{\sigma}\cdot{\bf q}$\ pion coupling, is {\it odd\/} under
spatial reflection of the pion, ${\bf q}\rightarrow -{\bf q}$.
It is thus responsible for the production of pions with odd angular
momentum ($p$-wave or higher).
Since, near threshold, the production of $p$-wave pions is suppressed
by the angular momentum barrier, we are interested here only in the
first term in Eq.~(\ref{eq:emit}), which, being {\it even\/} under
spatial reflection, can describe the production of $s$-wave pions.
The existence of this term is a consequence of Lorentz invariance,
which requires both a timelike and spacelike part to be present in the
original pseudovector ($\gamma_5\gamma_\mu q^\mu$) coupling.

Alternatively, coupling to an $s$-wave pion can also be generated from a
nonrelativistic $\bbox{\sigma}\cdot{\bf q}$\ coupling term by
invoking Galilean invariance.
Clearly, the coupling should involve a dot product between
$\bbox{\sigma}$\ and the momentum of the pion {\it relative\/} to
the nucleon, rather than just the momentum ${\bf q}$\ of the pion.
However, since $m_\pi$\ is so much smaller than the mass $M$\ of the
nucleon, the relative momentum differs from ${\bf q}$\ by only a
small recoil term involving the nucleon momentum ${\bf p}$.
Thus we expect a $\bbox{\sigma}\cdot{\bf p}$\ coupling term from
the nucleon recoil, which can produce $s$-wave pions, that is smaller than
the $\bbox{\sigma}\cdot{\bf q}$\ term by a factor of $m_\pi/M$.
The nonrelativistic limit of the first term in Eq.~(\ref{eq:emit}),
\begin{equation}
\overline{U}_{s^\prime}(p^\prime)\gamma_5\gamma_0 q_0 U_s(p)
\simeq -{m_\pi\chi^\dagger_{s^\prime}\bbox{\sigma}
\cdot({\bf p}+{\bf p}^\prime)\chi_s\over 2 M}\ ,
\label{eq:nonrel}
\end{equation}
is precisely this recoil contribution, where, near threshold, we take
$q_0\simeq m_\pi$.

It is straightforward to evaluate the contribution of the one-body
term in Fig.~\ref{fig1}(a) for the partial wave
$^3P_0\rightarrow{}^1S_0$.
The result for the corresponding matrix element is \cite{KOL66}
\begin{equation}
J_1=-{m_\pi^2\over pp^\prime}\int_0^\infty{\rm d}r\,
j_0\left(q^\prime r\over 2\right)
u_0(r)\left({{\rm d}\over{\rm d}r}+{1\over r}\right)u_1(r)\ .
\label{eq:one_body}
\end{equation}
Here $u_0$\ and $u_1$\ are distorted coordinate-space wave functions for the
$^1S_0$\ and $^3P_0$\ channels, $p$ and $p^\prime$ are the initial and final
relative momenta of the two nucleons, and $q^\prime$ is the pion momentum in
the final $pp$ center-of-mass frame.
The distorted waves are normalized such that
\begin{equation}
u_L(r)\rightarrow\sin\left(pr-{\pi L\over 2}+\delta_L\right)\ ,
\end{equation}
as $r\rightarrow\infty$, where $\delta_L$ is the corresponding phase
shift.
An analogous prescription applies when, asymptotically, Coulomb wave
functions are used.

Equation~(\ref{eq:one_body}) agrees with the result of
Ref.~\cite{KOL66}, except for the extra factor $j_0(q^\prime r/2)$, which
represents the $s$-wave pion wave function.
Near threshold, the pion momentum $q^\prime$\ is small and this factor is
close to unity.
However, the integral in Eq.~(\ref{eq:one_body}) involves only
scattering waves and no bound states.
It is thus very long ranged and is found to converge only if the pion
wave function is included.
In the work of Koltun and Reitan, the pion wave function was
neglected, and therefore it appears that their numerical result for
$J_1$\ (Table~II of Ref.~\cite{KOL66}) is in error.

When the IUCF Cooler data \cite{MEY92} became available, it was
immediately clear that the one-body cross section, calculated with
$J_1$\ alone, does {\it not\/} reproduce the data, underestimating the
experiment by about a factor of five \cite{MEY92}.
This discrepancy is so much larger than any uncertainty in the
calculation of $J_1$ that the conclusion is unavoidable that there
must be significant {\it additional\/} contributions to $s$-wave pion
production, over and above the one-body term discussed so far.


\subsection{Pion rescattering}
\label{sub:pion}

An obvious mechanism for pion production that involves both nucleons
is the so-called pion rescattering diagram shown in
Fig.~\ref{fig1}(b).
We note that in Eq.~(\ref{eq:one_body}) the wave function for a free
pion is used, while the two nucleons appear as distorted waves.
One might argue that the pion rescattering diagram represents the
first correction that arises from the distortion of the {\it pion}.
The matrix element $J_\pi$ for the rescattering contribution, derived
from a simple phenomenological pion-nucleon $s$-wave interaction, is
given in Ref.~\cite{KOL66} as
\begin{equation}
J_\pi={\lambda_1 m_\pi\over pp^\prime}\int_0^\infty{\rm d}r\,
j_0\left(q^\prime r\over 2\right)u_0(r)
\left[f(r)\left({{\rm d}\over{\rm d}r}+{1\over r}\right)
+{M\over m_\pi}\left(2+{m_\pi\over 2M}\right){{\rm d}f\over{\rm d}r}\right]
u_1(r)\ .
\label{eq:rescat}
\end{equation}
The radial function $f$ is defined as $f(r)\equiv e^{-\mu r}/r$, where
$\mu=\sqrt{3\over 4} m_\pi$.

The pion-rescattering matrix element in Eq.~(\ref{eq:rescat}) scales with the
parameter $\lambda_1$.
This parameter is obtained from the appropriate isospin average of the
pion-nucleon $s$-wave scattering lengths $a_{1/2}$\ and $a_{3/2}$,
corresponding to isospin $1\over 2$ and $3\over 2$:
\begin{equation}
\lambda_1=-{m_\pi\over 6}(a_{1/2}+2a_{3/2})\ .
\label{eq:lambda_def}
\end{equation}
Koltun and Reitan \cite{KOL66}, as well as other authors more recently
\cite{MIL91,NIS92}, use the older value of
\begin{equation}
\lambda_1^{\rm KR}=0.005\ ,
\label{eq:lambdaKR}
\end{equation}
while Lee and Riska \cite{LEE93}, making use of the computer code
SAID \cite{ARN91}, favor the value
\begin{equation}
\lambda_1^{\rm LR}=-0.0023\ .
\end{equation}
In both cases, pion-nucleon phase shifts have been extrapolated
down to threshold from energies where scattering data are available.
The two values differ because of new data added during the past 25
years and also because of different constraints in the extrapolation
procedure.
On the other hand, there exists experimental information that is more
directly related to the scattering lengths.
For instance, from the branching ratios of the decay of pionic
hydrogen \cite{SPU77}, one obtains $a_{1/2}-a_{3/2}=0.263\pm 0.005 m_\pi^{-1}$;
from the measured $1s$ width of pionic hydrogen \cite{BEE91}, one
obtains $2a_{1/2}+a_{3/2}=0.258\pm 0.012 m_\pi^{-1}$.
Combining the two results yields
\begin{equation}
\lambda_1=0.001\pm 0.002\ .
\label{eq:lambda}
\end{equation}
We conclude that so far there is no experimental evidence that
$\lambda_1$\ differs from zero or, consequently, that pion
rescattering contributes to $pp\rightarrow pp\pi^0$.
Nevertheless, we will explore the sensitivity of the calculated cross section
to
a possible nonzero value of $\lambda_1$ in Sec.~\ref{sec:discussion}.


\subsection{Exchange of heavy mesons}
\label{sub:exchange}

In this subsection we discuss the exchange of mesons heavier than the pion.
Clearly, meson exchange, where the intermediate state is a
positive-energy nucleon, is already contained in the distorted waves
used for evaluating the one-body term in Fig.~\ref{fig1}(a), as is
explicitly manifest in the construction of the Bonn potential.
However, a virtual, negative-energy state as shown in
Fig.~\ref{fig1}(c) is {\it not\/} contained in the nonrelativistic
wave functions.
Therefore, this contribution must be explicitly added to the pion
production operator as a two-body correction.

The matrix element for Fig.~\ref{fig1}(c) is calculated using
elementary Feynman rules \cite{BJO64}.
Let us first consider scalar-isoscalar $\sigma$-meson exchange
(other heavy mesons will be discussed below).
Neglecting distortions for a moment, this matrix element is given schematically
by
\begin{equation}
{\cal M}_{fi}\propto\overline{U}^\prime_2 U_2
\left(g_\sigma^2\over m_\sigma^2-k_\mu^2\right)\overline{U}^\prime_1
\left[\left(V\overline{V}\over 2 M\right)\gamma_5\gamma_0 q_0
+\gamma_5\gamma_0 q_0\left(V\overline{V}\over 2 M\right)\right]U_1\ ,
\label{eq:mea}
\end{equation}
where $k_\mu$ is the momentum of the exchanged $\sigma$ meson,
and $U_1$ and $U^\prime_1$ ($U_2$ and $U^\prime_2$)
are the initial and final Dirac spinors for the first (second) proton.
In Eq.~(\ref{eq:mea}) we have only included the antinucleon contribution
$V\overline{V}/2M$\ for the Feynman propagator of the intermediate nucleon,
taking the nonrelativistic limit $k_0+E_k\simeq 2M$ for the energy denominator
(there is an implicit sum over the spin indices of $V$ and $\overline{V}$).
Also, we have only kept the $\gamma_5\gamma_0 q_0$\ part of the
pion-nucleon vertex, since we are interested in $s$-wave pions.
The two terms in Eq.~(\ref{eq:mea}) correspond to the two possible
orderings for the emission of the $\sigma$ meson and the
emission of the pion.

It is a simple matter to take the nonrelativistic limit of
Eq.~(\ref{eq:mea}) by expanding to lowest order in $1/M$.
This yields
\begin{equation}
{\cal M}_{fi}\propto -{1\over M}
\left({g_\sigma^2\over m_\sigma^2+{\bf k}^2}\right)
{m_\pi\chi_1^{\prime\dagger}\bbox{\sigma}\cdot({\bf p}+{\bf p}^\prime)\chi_1
\over 2M}\ .
\label{eq:meb}
\end{equation}
Here ${\bf k}$\ is the momentum transfered by the $\sigma$ meson and
${\bf p}$\ and ${\bf p}^\prime$\ are the initial and final momenta
of the first nucleon.
Equation~(\ref{eq:meb}) then represents an additional two-body
contribution to the effective nonrelativistic operator that describes
pion production.
This term can be directly compared to the one-body term in
Eq.~(\ref{eq:nonrel}), which has a very similar form, except that the
two-body term contains an additional factor of $1/M$\ and a factor
$g_\sigma^2/({\bf k}^2+m_\sigma^2)$ from the $\sigma$-meson
propagator.
It is important to realize that the one-body and $\sigma$-meson
exchange contributions [Eqs.~(\ref{eq:nonrel}) and (\ref{eq:meb})]
have the same sign.
Thus the two-body contribution will interfere constructively with the
one-body term.

We now express Eq.~(\ref{eq:meb}) in coordinate space by taking the
Fourier transform, and then calculate the matrix element with
nonrelativistic distorted waves.
The momenta ${\bf p}$\ and ${\bf p}^\prime$\ become gradient operators that act
on the distorted waves.
These give rise to a factor $({\rm d}/{\rm d}r+1/r)$\ as in
Eq.~(\ref{eq:one_body}).
The meson propagator is transformed to a radial function $f_\sigma(r)$
that contains the mass $m_\sigma$ and the coupling constant
$g_\sigma$.
This function has the general form
\begin{equation}
f_x(r)={g_x^2\over 4\pi}{e^{-m_xr}\over r}\ ,
\label{eq:yukawa}
\end{equation}
where $x=\{\sigma,\delta,\omega,\rho\}$.
Finally, the contribution of Fig.~\ref{fig1}(c) from the exchange of
the $\sigma$ meson ($x=\sigma$) to pion production becomes
\begin{equation}
J_\sigma=-{m_\pi^2\over pp^\prime}\int_0^\infty{\rm d}r\,
j_0\left(q^\prime r\over 2\right)
u_0(r){f_\sigma(r)\over M}\left({{\rm d}\over{\rm d}r}+{1\over r}\right)
u_1(r)\ .
\label{eq:sigma}
\end{equation}
Equation~(\ref{eq:sigma}) is identical to Eq.~(\ref{eq:one_body})
except for the extra factor of $f_\sigma(r)/M$.
In contrast to the situation with $J_1$ [Eq.~(\ref{eq:one_body})], the
inclusion of the pion wave function $j_0(q^\prime r/2)$ is not crucial in this
case because of the short range of $f_\sigma(r)$.

As mentioned later, we obtain the $NN$ distorted waves from the
coordinate-space version of the Bonn one-boson-exchange potential.
The boson masses and coupling constants needed to construct this
potential have been fit to $NN$ scattering data and are listed in
Table~A.3 of Ref.~\cite{MAC89}.
For consistency, we use the {\it same\/} parameters in calculating the
exchange contributions discussed here.
We also adopt the technique described in Eq.~(A.28) of
Ref.~\cite{MAC89} to include monopole form factors at all
meson-nucleon vertices, according to the prescription
\begin{equation}
g_x\rightarrow g_x{\Lambda_x^2-m_x^2\over\Lambda_x^2-k_\mu^2}\ ,
\label{eq:form_factor}
\end{equation}
where $k_\mu$\ is the transferred momentum and $\Lambda_x$ is the
cutoff mass (also listed in Table~A.3 of Ref.~\cite{MAC89}).

In the following, we also consider the contributions to the diagram in
Fig.~\ref{fig1}(c) from the exchange of mesons other than the $\sigma$
meson, again using the corresponding parameters from the Bonn
potential.
We will find later that these contributions are small compared to
$J_\sigma$.

Let us begin with the scalar-isovector $\delta$ meson.
Its contribution $J_\delta$\ has the form of Eq.~(\ref{eq:sigma}) with
the appropriate mass $m_\delta$\ and coupling $g_\delta$\ used in
Eq.~(\ref{eq:yukawa}):
\begin{equation}
J_\delta=-{m_\pi^2\over pp^\prime}\int_0^\infty{\rm d}r\,
j_0\left(q^\prime r\over 2\right)
u_0(r){f_\delta(r)\over M}\left({{\rm d}\over{\rm d}r}+{1\over r}\right)
u_1(r)\ .
\label{eq:delta}
\end{equation}
For $pp\rightarrow pp\pi^0$\ the isospin factors are the same for the
exchange of either an isoscalar or an isovector meson.
The contribution from the exchange of a vector-isoscalar $\omega$
meson can be calculated in a similar fashion.
It contributes a term
\begin{equation}
J_\omega=-{m_\pi^2\over pp^\prime}\int_0^\infty{\rm d}r\,
j_0\left(q^\prime r\over 2\right)u_0(r)\left[{f_\omega(r)\over M}
\left({{\rm d}\over{\rm d}r}+{1\over r}\right)
+\left(1\over 2M\right){{\rm d}f_\omega\over{\rm d}r}\right]u_1(r)\ .
\label{eq:omega}
\end{equation}
The contribution from the exchange of a vector-isovector $\rho$ meson
has the form of Eq.~(\ref{eq:omega}) with the exception of a nonvanishing
tensor
coupling:
\begin{equation}
J_\rho=-{m_\pi^2\over pp^\prime}\int_0^\infty{\rm d}r\,
j_0\left(q^\prime r\over 2\right)u_0(r)\left[{f_\rho(r)\over M}
\left({{\rm d}\over{\rm d}r}+{1\over r}\right)
+\left(1+C_\rho\over 2M\right){{\rm d}f_\rho\over{\rm d}r}\right]u_1(r)\ .
\label{eq:rho}
\end{equation}
Here $C_\rho=6.1$\ is the ratio of the tensor to vector coupling
for the $\rho$ meson \cite{MAC89}.
Note that Eq.~(\ref{eq:rho}) describes only the $\rho$-meson contribution to
the diagram of Fig.~\ref{fig1}(c).
In principle, there are other contributions of the $\rho$ meson to the axial
charge that arise from a $\pi\rho$ current \cite{TOW92}.
However, the $\pi\rho$ current has an isospin factor proportional to
the dot product between the pion field and
$\vec{\tau}_1\times\vec{\tau}_2$, and therefore does not contribute to
the reaction $pp\rightarrow pp\pi^0$.
Again, we modify Eqs.~(\ref{eq:delta})--(\ref{eq:rho}) to include form factors
following Eq.~(A.28) of Ref.~\cite{MAC89}.

In principle, contributions from the exchange of pseudoscalar $\pi$
and $\eta$ mesons to the diagram in Fig.~\ref{fig1}(c) are possible.
These contributions would be large if pseudoscalar coupling ($\gamma_5$)
were used in Eq.~(\ref{eq:Lint}).
However, there are several arguments in favor of pseudovector coupling
($\gamma_5\gamma_\mu q^\mu$) \cite{HOR83,SER86}.
For example, pseudoscalar coupling implies an unrealistically large $s$-wave
pion-nucleon scattering length.
In contrast, pseudovector coupling yields zero (to lowest order) for
the scattering lengths, in agreement with experiment and consistent
with the (experimentally supported) value of $\lambda_1=0$\ for the
pion rescattering term (see Sec.~\ref{sub:pion}).
After adopting pseudovector coupling, MEC
contributions from $\pi$ and $\eta$ mesons vanish to lowest order in $1/M$.


\subsection{Calculational details and total cross section}
\label{sub:calculational}

The first step in the calculation is the evaluation of the $NN$
distorted waves $u_1$\ and $u_0$\ in the entrance and exit channels.
To this aim we make use of the coordinate-space version of the Bonn
one-boson-exchange potential, as described in Appendix~A.3 of
Ref.~\cite{MAC89}.
The form factor correction is applied according to the prescription in
Eq.~(A.28) of Ref.~\cite{MAC89}, and the Darwin term (proportional to
$\nabla^2$) is included following Ref.~\cite{BRY69}.
Boson masses, coupling constants, and cutoff masses are listed in
Table~A.3 of Ref.~\cite{MAC89} for all contributing mesons.
In that table, two sets of parameters are listed [Bonn potential A
(BPA) and Bonn potential B (BPB)]; they differ in the contribution of
the $\eta$ meson and in the value of the cutoff parameters.
In order to test for the sensitivity of the calculation to the
distorting potential and to connect to earlier results \cite{MEY92},
we have obtained distorted waves also from the Reid soft-core (RSC)
and Reid hard-core (RHC) potentials \cite{REI68}.

The next step is the evaluation of the one-body contribution  $J_1$\
[Eq.~(\ref{eq:one_body})].
This depends only on the $NN$ wave functions and the pion-nucleon
coupling constant.
For the latter, we adopt the value $f_{\pi NN}^2/4\pi=0.075$ which is
consistent with the Nijmegen phase-shift analysis \cite{BER90,STO93}.
In order to compute the very long-ranged integral in
Eq.~(\ref{eq:one_body}), we integrate conventionally from 0 to some
$r_{\rm max}$\ and then rotate the contour into the complex plane as
explained in Ref.~\cite{VIN70}.
This changes an integrand that is oscillating like $(\sin{r})/r$\ into
one that is exponentially damped.
For calculations without the Coulomb interaction, we choose $r_{\rm max}$\ to
be about $7\nobreak\,\text{fm}$.
When the Coulomb interaction is included, $r_{\rm max}$\ is increased
to $100\nobreak\,\text{fm}$ in order to be able to use the simplest asymptotic
form \cite{ABR70} for Coulomb wave functions in the complex plane.
Our results are insensitive to the exact choice of $r_{\rm max}$.
Since the initial state has a relatively large momemtum
($p\simeq 1.9\nobreak\,\text{fm}^{-1}$),
we expect that Coulomb effects in the entrance channel are small,
even at threshold.
Thus, in our calculation, the Coulomb interaction only affects the
$^1S_0$\ final state.

The next step, in principle, is determining the pion rescattering contribution
$J_\pi$, as in Eq.~(\ref{eq:rescat}).
However, since the rescattering parameter $\lambda_1$\ seems to be
consistent with zero, this step will be omitted except when we
investigate the sensitivity of the results to a possible deviation of
$\lambda_1$ from zero.

Finally, heavy-meson exchange contributions for $\sigma$, $\delta$,
$\omega$, and $\rho$\ mesons are calculated using
Eqs.~(\ref{eq:sigma}) and (\ref{eq:delta})--(\ref{eq:rho}).
These depend on meson coupling constants and masses and form-factor
cutoff masses.
For these parameters we use the same values that define the distorting
potentials BPA and BPB, discussed above.
For this reason, our calculations with BPA and BPB distortions are
{\it self consistent}.
Note that the calculation does not contain any parameters that are
adjusted to pion-production data.

The matrix element $J_{\rm tot}$ for the reaction $pp\rightarrow
pp\pi^0$\ is then composed of contributions from the one-body term
$J_1$, from pion rescattering $J_\pi$ (in principle), and from
heavy-meson exchange currents $J_{\rm MEC}$:
\begin{equation}
J_{\rm tot}=J_1+J_\pi+J_{\rm MEC}\ .
\label{eq:Jtot}
\end{equation}
As described in Ref.~\cite{KOL66}, the total cross section is obtained
as a phase space integral over the square of a matrix element,
\begin{equation}
\sigma_{\rm tot}={4f_{\pi NN}^2\over\pi\beta M m_\pi^5}
\int_0^{q^\prime_{\rm max}}{\rm d}q^\prime\, q^{\prime 2}
p^\prime|J_{\rm tot}|^2\ ,
\end{equation}
where $\beta$ is the lab velocity of the projectile, $q^\prime$ is the pion
momentum in
the final $pp$ center-of-mass system, and $p^\prime$ is the relative momentum
of
the final protons.


\section{DISCUSSION OF RESULTS AND SENSITIVITY TO INGREDIENTS}
\label{sec:discussion}


\subsection{Contributions to the matrix element}
\label{sub:contributions}

We now examine the relative importance of the various contributions to
the matrix element $J_{\rm tot}$\ [see Eq.~(\ref{eq:Jtot})].
To this aim, we evaluate $J_{\rm tot}$\ for the typical values of
$p=1.9\nobreak\,\text{fm}^{-1}$, $p^\prime=0.2\nobreak\,\text{fm}^{-1}$
and $q^\prime=0.1\nobreak\,\text{fm}^{-1}$.
The results are listed in Table~\ref{table1}, which shows that the
$\sigma$-meson contribution $J_\sigma$\ is about as large as the
one-body term $J_1$.
Furthermore, these two contributions are constructive; therefore, the
$\sigma$-meson contribution will increase the cross section by about a
factor of four!

The next important contribution is from the $\omega$\ meson (25--35\%
of the $J_1$\ or $J_\sigma$\ contribution).
In comparison, all other terms are significantly smaller. In
particular, the pion rescattering term $J_\pi$\ is small even when the
large, older value of
$\lambda_1=0.005$\ is used, and the isovector $\delta$- and
$\rho$-meson contributions are very small and tend to cancel each
other.
For the sake of completeness, we include in $J_{\rm MEC}$\ the
contributions from {\it all\/} heavy mesons; however, it is important
to keep in mind that the dominating two-body contribution is due to
the scalar $\sigma$-meson exchange.

This large $J_\sigma$ contribution [Eq.(\ref{eq:sigma})] can be easily
understood in relativistic models.
In these models \cite{SHE83,CLA83,SER86}, the nucleon has an effective
mass
\begin{equation}
M^\ast=M-S
\end{equation}
from a strong repulsive scalar mean field $S$.
This enhances the lower components in the Dirac spinors $U$.
If spinors of mass $M^\ast$\ are used, Eq.~(\ref{eq:nonrel}) becomes
\begin{equation}
\overline{U}_{s^\prime}(p^\prime;M^\ast)\gamma_5\gamma_0 q_0 U_s(p;M^\ast)
\simeq
-{m_\pi\chi_{s^\prime}^{\prime\dagger}\bbox{\sigma}\cdot({\bf p}+{\bf
p}^\prime)
\chi_s\over 2M^\ast}\ .
\end{equation}
Thus a reduction of the nucleon mass {\it enhances\/} the $s$-wave
pion coupling.
Let us expand $1/M^\ast$\ to lowest order in $1/M$:
\begin{equation}
{1\over M^\ast}\simeq{1\over M}\left(1+{S\over M}\right)\ .
\end{equation}
If one assumes that the scalar field arises from the second nucleon,
or $S=f_\sigma(r)$\ [see Eq.~(\ref{eq:yukawa})], then the second term
($S/M$), when substituted into Eq.~(\ref{eq:one_body}), immediately
yields the $\sigma$-meson contribution of Eq.~(\ref{eq:sigma}).
Thus the scalar field of the second nucleon affects the Dirac spinor
of the first nucleon in such a way that the modified spinor has a
larger coupling to an $s$-wave pion.
This simple argument also shows that $J_1$\ and $J_\sigma$\ should add
constructively.

It is nevertheless surprising that the $\sigma$\ contribution is
almost as large as the one-body term, especially since $J_\sigma$\
involves an additional factor of $1/M$.
This is explained by realizing that it is not $J_\sigma$\ that is
large, but $J_1$\ that is anomalously small.
Near threshold, the two nucleons approach each other with a relative
momentum $p\simeq 1.9\nobreak\,\text{fm}^{-1}$, and yet they must almost stop
in order to produce a pion.
Because it is difficult to mediate such a large momentum mismatch
through the distortions of Fig.~\ref{fig1}(a), $J_1$\ is relatively
small.
On the other hand, the $\sigma$-meson propagator in Fig.~\ref{fig1}(c)
provides an efficient means of transferring momentum.
As already pointed out, it is interesting that the $\sigma$-meson mass
($550\nobreak\,\text{MeV}$, according to Table~A.3 of Ref.~\cite{MAC89}) is
comparable to $p\simeq 1.9\nobreak\,\text{fm}^{-1}$.

The interpretation of the $\sigma$-meson contribution deserves comment.
Clearly the $\sigma$ is not a sharp resonance.
Instead it is a simple phenomenological model for the important
intermediate-range attraction in the $NN$ interaction.
Assuming that this attraction transforms as a Lorentz scalar provides a natural
explanation of the spin dependence of the $NN$ force \cite{SHE83,CLA83}.
We expect a more complicated model of the intermediate-range attraction (such
as
correlated two-pion exchange) to yield a similar $J_\sigma$ contribution
provided the attraction transforms as a Lorentz scalar.


\subsection{Total cross section results}
\label{sub:total}

In Fig.~\ref{fig2} our calculation is compared with the available
$pp\rightarrow pp\pi^0$\ total cross section data as a function of $\eta$,
the maximum pion momentum in the overall center-of-mass frame in units of
$m_\pi$ ($\eta\equiv q_{\rm max}/m_\pi$), or,
alternatively, the projectile energy $T$ in the laboratory.
The IUCF Cooler data \cite{MEY92} are shown as solid dots
(note that there is a 6.6\% uncertainty in the normalization of the data
that is not shown),
while data from previous work are marked with crosses \cite{STA58}, squares
\cite{DUN59}, bars \cite{SHI82}, and diamonds \cite{STA90}.
As is well known by now, the one-body term alone (dashed line) greatly
underestimates the data.
However, when the two-body contributions are included (solid line),
the measured cross sections are reproduced to an extent that is truly
remarkable in view of the fact that none of the parameters of the
model have been adjusted to pion-production data.
The dot-dashed line, which has been obtained without the Coulomb
interaction, demonstrates that Coulomb repulsion is responsible for a
fairly sizeable reduction of the cross section near threshold.
All calculations shown in Fig.~\ref{fig2} use the BPA distorting potential.

It has been pointed out earlier \cite{MEY92} that the energy
dependence of the $s$-wave cross section follows from phase space and
the final-state interaction between the two (charged) protons.
This is sufficient to reproduce the shape of the measured cross
section up to $\eta\simeq 0.6$, where higher partial waves (which we
are neglecting) start to contribute \cite{MEY92}.
Thus it is only the {\it magnitude\/} of the cross section near
threshold, represented by a single number, that contains nontrivial
physics information.
Our work shows that heavier meson exchange (mainly of the $\sigma$ meson),
together with the one-body term, is sufficient to explain the observed
magnitude
of the cross section.
Furthermore, there are no alternative explanations at this time.
In the past, it has been suspected that the role of heavy-meson exchange
currents is suppressed by $NN$ correlations and form factors at the
meson-nucleon vertices.
We find that this is not the case in the $NN$ system:
Because of the simplicity of the present reaction, it is possible to
explicitly include $NN$ correlations by solving for the full two-body
wave function.
Even with these correlations included, $J_\sigma$ remains large.

The dependence of $J_\sigma$\ on the cutoff mass is illustrated in
Fig.~\ref{fig3} (using BPA).
Note that the form factor in Eq.~(\ref{eq:form_factor}) has a
normalization at $k_\mu^2=0$ that depends strongly on
$\Lambda_\sigma$:
\begin{equation}
g_\sigma(0)
=g_\sigma{\Lambda_\sigma^2-m_\sigma^2\over\Lambda_\sigma^2}\ .
\label{eq:gzero}
\end{equation}
Indeed, the primary effect of this form factor is to change the value
of $g_\sigma(0)$\ rather than the momentum dependence of the
interaction.
It is this coupling near $k_\mu^2=0$\ that plays a dominant role when
adjusting the parameters of one-boson-exchange potentials to $NN$
scattering.
Therefore it may be more meaningful to compare results with different
$\Lambda_\sigma$\ values at a fixed $g_\sigma(0)$\ rather than at a
fixed $g_\sigma$.
Figure~\ref{fig3} shows only a modest decrease of $J_\sigma$\ with
decreasing $\Lambda_\sigma$\ (at fixed $g_\sigma$).
However, if one keeps $g_\sigma(0)$\ fixed rather than $g_\sigma$,
$J_\sigma$\ actually increases very slightly with decreasing
$\Lambda_\sigma$.
If two potentials fit phase shifts with different cutoff masses, we expect
approximately similar values of $g_\sigma(0)$ (rather than $g_\sigma$).
If this is the case, the $J_\sigma$ contribution will be almost independent of
$\Lambda_\sigma$.

We now examine the sensitivity to the various distorting potentials
that are mentioned in Sec.~\ref{sub:calculational}.
Figure~\ref{fig4} shows the total cross section, calculated with the
Coulomb interaction included and without the pion rescattering term,
for the two somewhat different one-boson-exchange potentials BPA
(solid line) and BPB (dotted line), and for the RSC (dot-dashed line)
and RHC (dashed line) potentials.
For the phenomenological Reid potentials there is no way to
unambiguously determine the meson-exchange contributions.
Therefore, we simply adopt the meson couplings and cutoff masses from
the Bonn potential BPA for both, RSC and RHC.
This allows us to study the effects of a change in only the distorted
waves $u_0$\ and $u_1$.
The RHC wave function is identically zero at small distances; in
contrast, the RSC wave function is nonzero, but still small.
This enhancement of the wave function at small $r$ leads to a modest
increase in pion production.
At small $r$, the BPA and BPB wave functions are almost identical, but
larger still than the RSC wave function, resulting in a slightly
increased cross section.
However, some of the difference between the Bonn and Reid results is
due to the fact that BPA and BPB generate slightly larger on-shell
phase shifts as compared to the Reid potentials.
This is because the Bonn potentials were fit to $pn$ rather than $pp$ data.
Nevertheless, the range in cross section for different potentials is
still relatively modest.
We point out that {\it all\/} calculations {\it without\/} meson exchange
(the lower four curves in Fig.~\ref{fig4}) greatly underestimate the data.

Based on the small value for $\lambda_1$ in Eq.~(\ref{eq:lambda}), we
do not expect pion rescattering to be important.
For completeness sake, we still wish to examine the effect of a
possible rescattering contribution.
All calculations in Fig.~\ref{fig5} are carried out with the BPA and
with the Coulomb interaction included.
The dotted lines correspond to $\lambda_1=0.005$, the dot-dashed lines
to $\lambda_1=-0.0023$, and the solid line is without rescattering.
The effect from pion rescattering is relatively small even for the
large value $\lambda_1=0.005$, and all calculations with only one-body
and pion-rescattering contributions (the lower two curves in
Fig.~\ref{fig5}) fall substantially below the experimental data.
Clearly pion rescattering as described by $J_\pi$\ [see Eq.~(\ref{eq:rescat})]
cannot explain the difference between $J_1$ and the data.
However, one must realize that Eq.~(\ref{eq:rescat}) is based on a simple
on-shell model for pion-nucleon scattering.
It is conceivable that the pion-nucleon interaction might be modified
due to the fact that the intermediate pion in Fig.~\ref{fig1}(b) is
off the mass shell.
This is a topic that deserves further study.


\section{IMPLICATIONS AND CONCLUSIONS}
\label{sec:implications}

We have calculated $s$-wave pion production in
$pp\rightarrow pp\pi^0$\ by considering the one-body term, pion
rescattering, and two-body meson-exchange processes.
We confirm that the one-body term underestimates the data by about a
factor of five, and that pion rescattering is indeed small.
On the other hand, we find a large contribution from the exchange of
heavy mesons (in particular, the scalar $\sigma$ meson) coupling to
the negative-energy state of a nucleon, as in Fig.~\ref{fig1}(c).
This meson-exchange contribution (MEC) is large enough to explain the
discrepancy between one-body production and the data, and, when taken
into account with self-consistent distortions, leads to an excellent
fit to the data without parameters that are adjusted to
pion-production information.

The theoretical description of the $pp\rightarrow pp\pi^0$\ reaction
close to threshold is clean and simple.
Only a single partial wave is allowed in either entrance or exit
channel, pion rescattering is suppressed, and intermediate $\Delta$
isobars are expected to be unimportant (in marked contrast to
most other reactions involving pions).
Furthermore, the simplicity of the system allows a full treatment of
$NN$ correlations.

The operator for the production of $s$-wave pions has the same form as
the axial-charge operator.
Therefore, we conclude that the axial charge in nuclear systems is
much larger than one is lead to believe from one-body predictions.
This agrees with the conclusion from a number of calculations of
first-forbidden $\beta$ decays in nuclei \cite{TOW92}.
However, these nuclear studies are less conclusive because of
structure ambiguities.
The $pp\rightarrow pp\pi^0$\ system is free of such ambiguities.

The $pp\rightarrow pp\pi^0$\ reaction is sensitive to two-body
contributions because of the nature of the axial-charge operator and
due to a large momentum mismatch that naturally favors the exchange of
heavy mesons.

Uncertainties in the meson-nucleon form factors, the two-nucleon wave function,
and pion rescattering have been examined and are much smaller than the two-body
meson-exchange contributions.
We are thus led to the conclusion that the near-threshold
$pp\rightarrow pp\pi^0$\ data provide {\it direct experimental
evidence for meson-exchange contributions to the axial current}.
This is analogous to the electro-disintegration of the deuteron which
provides direct evidence for electromagnetic MEC.
However, in our system the MEC is a very large effect, compared to a $\sim
10\%$
contribution to the cross section in the case of the electromagnetic current
\cite{RIS72}.

Furthermore, near-threshold $pp\rightarrow pp\pi^0$\ data provide
direct evidence for exchange currents from {\it heavy\/} mesons.  To
the best of our knowledge, all previous experimental evidence for MEC
has involved exclusively pions.
This is significant because of the possibility that heavy-meson
contributions are greatly suppressed by $NN$ correlations and
meson-nucleon form factors.
We have shown that this is {\it not\/} the case in the $pp$ system.
The importance of heavy-meson exchange currents could be very
significant for experiments planned at CEBAF, which, involving higher
momentum transfer, are likely to be sensitive to such short-distance
effects.

Relativistic nuclear models characteristically feature large Lorentz
scalar and vector potentials \cite{CLA83,SER86}.
In these models, the $\sigma$\ meson describes the important
intermediate-range attraction in the $NN$ interaction and gives rise
to a scalar potential that reduces the effective mass of a nucleon.
This reduction in $M^\ast$\ enhances the lower components of Dirac
wave functions and this increases the axial charge.
This change in the Dirac wave functions is also what provides the
natural description of a large range of nucleon-nucleus scattering
data, in particular spin observables \cite{CLA83}.
The fact that we find that this same effect explains the observed pion
production cross section also provides an indirect experimental
confirmation of this key feature of relativistic models.

The $J_\sigma$\ term in our calculation can be viewed more generally as a term
involving some intermediate-range attraction in the $NN$ interaction that
transforms like a Lorentz scalar.
This contribution does not necessarily have to arise from an
elementary narrow $\sigma$\ meson.
Instead, it could well be an effective representation of a more
complex mechanism.
In any case, and whatever the microscopic origin of this Lorentz
scalar attraction, its importance in the present calculation provides
evidence for a large relativistic effect in the $NN$ interaction.

Future theoretical work on the $pp\rightarrow pp\pi^0$\ reaction
should be devoted to a study of the off-shell aspects of pion
rescattering.
Also, the present study should be extended to include the next higher
partial waves, as they become important with increasing bombarding
energy.
This is especially important, since a measurement of the
spin-dependent total cross section is planned at IUCF that will allow
the separation of $p$-wave pion contributions.
This provides the data necessary to search for meson-exchange
contributions to Gamow-Teller matrix elements.
Since we have found a large and unexpected meson-exchange contribution
to the axial charge, we may well speculate about the existence of a
similar contribution to Gamow-Teller matrix elements.

This could have important consequences for the
$pp\rightarrow d e^+\nu$\ reaction and the solar neutrino problem
\cite{BAH92}.
This reaction is believed to proceed via a $^1S_0\rightarrow{}^3S_1$\
Gamow-Teller transition.
An enhancement of as little as 15\% in the matrix element for
$pp\rightarrow d e^+\nu$\ (from an unexpected MEC) would dramatically
reduce the disagreement between theory and experiment \cite{CAS92},
because, in the standard solar model, the rate of this reaction
sensitively affects the central temperature of the sun and thus the
high-energy neutrino flux.
Of course, present calculations of $\pi$ and $\rho$ MEC \cite{CAR91}
give only a small contribution.
Furthermore, MEC for this channel are expected to be
smaller than the order $v/c$ MEC contributions to the axial charge.
However, one may still speculate that some {\it unsuspected\/}
MEC or other effect could be important.
Through very accurate pion production data it may be possible to gain
(indirect) experimental information about the
$pp\rightarrow d e^+\nu$\ rate.


\acknowledgements

C.J.H.\ and D.K.G.\ acknowledge support from the U.S.\ Department of Energy
under Grant No.\ DE--FG02--87ER--40365.
H.O.M.\ acknowledges support from the U.S.\ National Science Foundation under
Grant No.\ NSF--PHY--9015957.



\begin{figure}
\caption{Contributions to the reaction $pp\rightarrow pp\pi^0$.
Shown are (a) the one-body term, (b) the two-body term that arises from
pion rescattering, and (c) the two-body term that arises
from the exchange of heavier mesons ($x$).}
\label{fig1}
\end{figure}
\begin{figure}
\caption{Total cross section for $pp\rightarrow pp\pi^0$\ as a function of
$\eta$, the maximum pion momentum in the center-of-mass frame
in units of $m_\pi$, or the projectile energy $T$ in the lab frame.
The recent IUCF Cooler measurement \protect{\cite{MEY92}} is shown by solid
dots; data from previous work are marked with crosses \protect{\cite{STA58}},
squares \protect{\cite{DUN59}}, bars \protect{\cite{SHI82}}, and
diamonds \protect{\cite{STA90}}.
The solid line includes heavy-meson exchange and the Coulomb interaction,
and the dotted lines describe the error bands for
the $\pi N$ coupling constant $f_{\pi NN}^2/4\pi=0.075\pm 0.003$ and the
rescattering parameter $\lambda_1=0.000\pm 0.002$ (see text).
The dashed line neglects heavy-meson exchange and includes the Coulomb
interaction, while the dot-dashed line includes heavy-meson exchange but
neglects the Coulomb interaction.
(Note that the Coulomb interaction is included in all other curves in this
paper.)
All calculations use BPA.
The solid curve is the same as in Figs.~\protect{\ref{fig4}} and
\protect{\ref{fig5}}.}
\label{fig2}
\end{figure}
\begin{figure}
\caption{Dependence of $J_\sigma$\ on the cutoff mass $\Lambda_\sigma$ for an
initial nucleon momentum of $p=1.9\nobreak\,\text{fm}^{-1}$, a final nucleon
momentum of
$p^\prime=0.2\nobreak\,\text{fm}^{-1}$, and a pion momentum of
$q^\prime=0.1\nobreak\,\text{fm}^{-1}$.
The dashed line is obtained by keeping $g_\sigma$ fixed, while for the solid
line $g_\sigma(0)$\ was kept constant [see Eq.~(\protect{\ref{eq:gzero}})].
The calculations use BPA, which has the standard cutoff
$\Lambda_\sigma=2\nobreak\,\text{GeV}$.}
\label{fig3}
\end{figure}
\begin{figure}
\caption{Sensitivity of the calculated cross section to the distorting
potential.
Shown is the cross section divided by $\eta^2$ as a function of $\eta$.
The data are the same as in Fig.~\protect{\ref{fig2}}.
The curves are for the Bonn potentials A (solid lines) and B (dotted
line), and for the Reid soft-core (dot-dashed line) and Reid hard-core
(dashed line) potentials.
The solid curve is the same as in Figs.~\protect{\ref{fig2}} and
\protect{\ref{fig5}}.
The lower four curves are without $J_{\rm MEC}$ (the curves are labeled in the
same manner as above, except a dot-dot-dashed curve is used for BPA).}
\label{fig4}
\end{figure}
\begin{figure}
\caption{Sensitivity of the calculated cross section to the
rescattering contribution using BPA.
Shown is the cross section divided by $\eta^2$ as a function of $\eta$.
The data are the same as in Fig.~\protect{\ref{fig2}}.
The curves are for $\lambda_1=0.005$ (dotted line), $\lambda_1=0$
(solid line), and $\lambda_1=-0.0023$ (dot-dashed lines).
The solid curve is the same as in Figs.~\protect{\ref{fig2}} and
\protect{\ref{fig4}}.
The lower two curves are without $J_{\rm MEC}$.}
\label{fig5}
\end{figure}


\begin{table}
\caption{Contributions to the total matrix element $J_{\rm tot}$ for
an initial nucleon momentum of $p=1.9\nobreak\,\text{fm}^{-1}$, a final nucleon
momentum of $p^\prime=0.2\nobreak\,\text{fm}^{-1}$, and a pion momentum of
$q^\prime=0.1\nobreak\,\text{fm}^{-1}$.
The Bonn one-boson-exchange potentials (BPA and BPB) and the Reid
potentials (RSC and RHC) are described in the text.
The pion-rescattering contribution $J_\pi$ assumes $\lambda_1=0.005$
[see Eq.~(\protect{\ref{eq:lambdaKR}})].}
\begin{tabular}{lcccccc}
Potential&$J_1$&$J_\sigma$&$J_\omega$&$J_\pi$&$J_\delta$&$J_\rho$\\
\tableline
BPA & $-0.174$ & $-0.163$ & $-0.062$ & $-0.042$ & $-0.009$ & $0.009$\\
BPB & $-0.165$ & $-0.152$ & $-0.063$ & $-0.044$ & $-0.024$ & $0.011$\\
RSC & $-0.143$ & $-0.135$ & $-0.050$ & $-0.038$ & $-0.006$ & $0.008$\\
RHC & $-0.126$ & $-0.140$ & $-0.048$ & $-0.040$ & $-0.007$ & $0.009$\\
\end{tabular}
\label{table1}
\end{table}


\end{document}